\documentclass[amsmath,twocolumn,showpacs,amssymb,aps,nofootinbib]{revtex4-2}
\usepackage{geometry}
\usepackage{graphicx}% Include figure files
\usepackage{dcolumn}% Align table columns on decimal point
\usepackage{bm}% bold math
\usepackage{epsfig,amsmath}
\usepackage{amssymb}
\usepackage{subfigure,esint}
\usepackage{float}
\usepackage{comment}
\usepackage[usenames,dvipsnames]{xcolor}
\usepackage[pagebackref=false, colorlinks=true]{hyperref}
\definecolor{redish}{rgb}{0.7,0.2,0.0}  % color defined in (r=red,g=green,b=blue) model
\definecolor{bluish}{rgb}{0.2,0.5,0.8}

\hypersetup{linkcolor=redish,          % color of internal links
	citecolor=magenta,        % color of links to bibliography
	filecolor=magenta,      % color of file links
	urlcolor=teal}          % color of the url

\begin{document}
\author{Ankit Gill}\email{ankitgill20@iitk.ac.in}
\affiliation{
Department of Physics, Indian Institute of Technology Kanpur, \\ Kanpur 208016, India}
\author{Tapobrata Sarkar}\email{tapo@iitk.ac.in}
\affiliation{
	Department of Physics, Indian Institute of Technology Kanpur, \\ Kanpur 208016, India}
\title{Speed Limits and Scrambling in Krylov Space}

\begin{abstract}
We investigate the relationship between Krylov complexity and operator quantum speed limits (OQSLs) of the complexity operator and level repulsion in random/integrable matrices and many-body systems. An enhanced level-repulsion corresponds to increased OQSLs in random/integrable matrices. However, in many-body systems, the dynamics is more intricate due to the tensor product structure of the models. Initially, as the integrability-breaking parameter increases, the OQSL also increases, suggesting that breaking integrability allows for faster evolution of the complexity operator. At larger values of integrability-breaking, 
the OQSL decreases, suggesting a slowdown in the operator's evolution speed. Information-theoretic properties, such as scrambling, coherence and entanglement, of Krylov basis operators in many-body systems, are also investigated. The scrambling behaviour of these operators exhibits distinct patterns in integrable and chaotic cases. For systems exhibiting chaotic dynamics, the Krylov basis operators remain a reliable measure of these properties of the time-evolved operator at late times. However, in integrable systems, the Krylov operator's ability to capture the entanglement dynamics is less effective, especially during late times. 
\end{abstract}
\maketitle

\section{Introduction}

Quantum chaos refers to the study of the signatures of quantum systems that exhibit characteristics similar to those with chaotic classical limits. In the latter, slight differences in initial conditions can lead to vastly different dynamics, making long-term predictions difficult. One of the primary indicators of quantum chaos is the level spacing distribution ~\cite{Bohigas:1983er,berry:tabor}, 
which describes the statistical properties of the energy levels of a quantum system. Another key quantity is the Loschmidt echo \cite{Peres:1984pd}, which measures the sensitivity of 
a quantum system to perturbations, by comparing the time evolution of an initial state with and without a small perturbation. This quantity can provide insights into the stability and 
reversibility of quantum evolution. The chaotic behaviour in quantum systems can also be explored through other dynamical quantities like Out-of-Time-Ordered Correlators (OTOCs) 
~\cite{vonKeyserlingk:2017dyr, Maldacena:2015waa, kitalk} and by information theoretic quantities like entangling power etc.
%add two citations

Recently, significant attention has been directed toward studying the dynamics of operators in Heisenberg's picture, particularly in exploring quantum chaos \cite{Srednicki:1994mfb}. 
Even closed quantum systems can exhibit thermalisation behaviour, where the long-time behaviour of observables can be described by thermal ensembles 
\cite{Deutsch:1991msp,Rigol:2007juv,DAlessio:2015qtq}. Such investigations also include studying OTOCs, circuit complexity, and operator entanglement entropy. 
The OTOC is a powerful diagnostic of operator growth, which measures the spreading of a local operator $O_{1}$ by the correlation function 
$ \langle O^{\dagger}_{1}(t) O^{\dagger}_{2}(0) O^{\dagger}_{1}(t) O^{\dagger}_{2}(0) \rangle$ with another local operator $O_{2}$ . The OTOC has been extensively utilised to study 
the scrambling of quantum information. However, since scrambling does not necessarily imply chaos ~\cite{Swingle:2016var,Xu:2019lhc}, its utility is limited to systems 
exhibiting semi-classical or large-N limits ~\cite{Sachdev:1992fk,Maldacena:2016hyu}. 

Useful in this context is the notion of complexity, 
a concept that lies at the intersection of computer science, 
quantum computing, and black hole physics. An important and commonly studied notion in this context is the Circuit complexity \cite{Dowling:2006tnk} 
that quantifies the minimal number of 
elementary gates required to construct the target unitary operator. It is the size of the smallest circuit that produces the desired target state from a initial product state. This concept 
has been studied ~\cite{Bhattacharyya:2018bbv,Ali:2018fcz} to characterise operator dynamics. However, its computation is limited to only a 
few systems ~\cite{Jaiswal:2021tnt,Pal:2022rqq}, notably integrable ones, due to challenges in finding the optimal combination of gates for the shortest circuit in 
generic chaotic quantum systems\cite{Brandao:2019sgy}. The difficulty is that the gates lying in later layers of the circuit may cancel the gates in previous layers.

The authors of \cite{Parker:2018yvk} introduced another notion of operator complexity, namely ``Krylov complexity" ($K$-complexity), to characterise operator growth under Heisenberg evolution. The main idea underlying $K$-complexity is that due to time evolution, the operator $O(t)$ evolves into an increasingly complex non-local operator whose representation in any basis of local operators requires an exponentially large number of coefficients. Hence, it is easier to treat operators of similar ``complexity" as a thermodynamic bath and look at the dynamics of the operator as it flows through the baths of increasing ``complexity". Their approach is based on a well-known recursion method \cite{muller}, widely used for probing dynamical properties (correlation functions) of condensed matter systems in linear response theory. The recursion method allows for systematically constructing an orthogonal basis (Krylov basis) of operators under the Heisenberg time evolution. $K$-complexity measures how an operator $O(t)$ grows over time under Heisenberg evolution $e^{-i \mathcal{H} t}$ in a basis which is fixed by the operator $O(0)$ and 
the Hamiltonian $\mathcal{H}$. A $K$-complexity calculation can be mapped to the problem of finding the average position of a quantum particle on a half-chain, with hopping matrix elements given by the Lanczos coefficients $b_n$. The  
``operator growth hypothesis" states that $b_n$ grows as fast as possible (linearly) in chaotic quantum systems.

Recently, there has been a flurry of works \cite{Baggioli:2024wbz,Nandy:2024htc,Chen:2024imd,Menzler:2024atb,Sasaki:2024puk,Bhattacharya:2024uxx,Basu:2024tgg,Craps:2023ivc,Balasubramanian:2022tpr,Camargo:2023eev,Caputa:2024vrn,Das:2024zuu,Alishahiha:2024rwm,Adhikari:2022oxr} on notions similar to the $K$-complexity. It also has been extended to integrability-chaotic crossovers ~\cite{Bhattacharya:2023zqt,Camargo:2024deu,Balasubramanian:2024ghv}, periodically driven systems \cite{Nizami:2023dkf}, field theory ~\cite{Avdoshkin:2022xuw,Adhikari:2022whf,Vasli:2023syq}, scrambling ~\cite{Bhattacharjee:2022vlt,Huh:2023jxt,Das:2020nay}, open quantum systems \cite{Bhattacharya:2022gbz,Bhattacharjee:2022lzy,Carolan:2024wov}, random matrix models  ~\cite{Bhattacharyya:2023grv,Bhattacharjee:2024yxj}, random walks \cite{Jeevanesan:2023ogo}, and adiabatic gauge potentials ~\cite{Bhattacharjee:2023dik,Takahashi:2023nkt}.

In addition to complexity, quantum speed limits ~\cite{Deffner:2017cxz,Margolus:1997ih,Srivastav:2024apk,Mohan:2021rel} have also been employed to characterise operator dynamics. Quantum speed limits are fundamental bounds on the minimum time required for a quantum system to evolve from one state to another. Traditional quantum speed limits tend to be overly conservative when estimating relevant timescales for various processes, such as thermalisation \cite{Eisert:2014jea}. Notably, the pioneering work \cite{Mand1945} spurred the development of more tailored speed limits for observables. Operator quantum speed limits (OQSLs) establish fundamental bounds on the rate at which quantum operators can evolve, for example, time evolution in the Heisenberg's picture. These limits are essential for understanding the maximum speed of quantum information processing \cite{Lloyd:2000cry} and thermalisation in many-body systems \cite{2020NatPh..16.1211N}. In \cite{Carabba:2022itd}, the authors generalised quantum speed limits for unitary operator flows by quantifying distances over the unitary flow. It has been used to constrain the linear dynamical response of quantum systems and the
quantum Fisher information, a central quantity in quantum metrology.  

In view of the above background, the purpose of this paper is to study aspects of operator dynamics in Krylov space and probe its relationship
with quantum chaos. In particular, we study
$K$-complexity and OQSLs for random and many body Hamiltonians, exemplified by a random matrix theory and the axial next-nearest neighbour 
Ising (ANNNI) model, respectively. In order to gain further insights into Krylov-basis operators, we analyse two specific information-theoretic tools,
namely entanglement and scrambling in Krylov space. These strengthen our understanding of the dynamics of chaotic quantum systems.

In the next section, we will demonstrate that $K$-complexity can also be computed as the expectation value of the ``Complexity Operator" $\mathcal{K}(t)$ in the Heisenberg picture \cite{Caputa:2021sib}. This approach has garnered significant attention in the field of OQSLs ~\cite{Carabba:2022itd,Bhattacharya:2024uxx,Hornedal:2023xpa,Hornedal:2022pkc}. Additionally, it is of considerable interest to study the OQSL of this operator to gain deeper insights into the dynamics of quantum complexity.

We further recall that in \cite{Hornedal:2022pkc}, a fundamental bound on the growth rate of Krylov complexity was derived by analytically exploring the conditions for its saturation. In ~\cite{Carabba:2022itd,Hornedal:2023xpa}, an OQSL was used to study the speed limit of the Complexity operator. The OQSL of the complexity operator is saturated when the so-called ``complexity algebra" is closed. In this work, we numerically investigate the OQSL of the complexity operator in both random/integrable matrices \cite{Bohigas:1983er} and many-body systems. Our analysis encompasses integrable and chaotic regimes, allowing us to compare the behaviour of the OQSL of $\mathcal{K}(t)$ across different types of quantum systems. The energy levels of integrable and chaotic quantum systems follow Poisson and Wigner-Dyson level spacing distributions respectively \cite{shastry:imt}. The advantage of studying the random/integrable matrices is to study the $K$-complexity and OQSL in cases where there is no notion of tensor product structure. Hence, only the level statistics determine the behaviour of $K$-complexity and OQSL. We will also discuss the impact of integrability-breaking in qubit Hamiltonian on OQSLs and the properties of Krylov basis operators as their complexity increases, which also provides information about the complexity of the thermodynamic baths of similar complexity through which $O(t)$ evolves. This discussion also aims to elucidate the relationship between operator complexity and other information-theoretic aspects of operator dynamics, like scrambling and entanglement entropy \cite{2002PhRvA..66d4303W}. Scrambling  ~\cite{Garcia:2022abt,Hayden:2007cs,Shenker:2013pqa} refers to the process by which quantum information becomes distributed across the degrees of freedom in a system, making it inaccessible to local measurements. By examining the average size of Krylov basis operators, we gain insights into how widely an operator spreads over the system's basis states. Coherence measures the superposition of these basis states, providing a quantitative understanding of the operator's complexity. For this purpose, 
we study the Krylov basis operators' average size, coherence, and entanglement properties generated by the Lanczos algorithm in many-body systems. These signatures also differentiate the scrambling mechanisms of Krylov basis operators. We also verify if these properties exhibited by Krylov basis operators are comparable to those of the time-evolving operator. In summary, this paper aims to provide a
comprehensive picture of operator dynamics in many-body systems by analysing Krylov basis operators' average size, coherence, and entanglement.

This paper is organised as follows. In the next sections \ref{sec:headings} and \ref{secthree}, we introduce the necessary notations and conventions
to be used in the rest of the paper. In section \ref{secfour}, we present our results on $K$-complexity and OQSL for Random Matrix theory and Many-body Hamiltonians.
Section \ref{secfive} contains our results on entanglement and scrambling in ANNNI model, and finally section \ref{secsix} concludes the paper
with a summary of the results. This paper also contains an appendix where we present a few details of some relevant calculations.

\section{$K$-Complexity and Operator Speed Limits}
\label{sec:headings}

In this section, we describe the recursion method \cite{muller} and the definition of $K$-complexity for an operator $O$ evolving under the time evolution generated by Hamiltonian $\mathcal{H}$. The Lanczos algorithm constructs the so-called Krylov basis. In this basis, the Liouvillian $\mathcal{L} \equiv [\mathcal{H},.]$ takes a tridiagonal form in case of  hermitian operator. Time evolution of an operator in Heisenberg picture is
\begin{equation}
\label{te_op}
O(t) = e^{i \mathcal{H} t} O e^{-i \mathcal{H} t}.
\end{equation}
The Krylov space is the minimal subspace where the dynamics of $O$ takes place. This subspace structure is evident from the power series expansion of Eq. \eqref{te_op},
\begin{equation}
\label{pos}
O(t) = O + \sum_{n} \frac{(it)^{n}}{n!} \mathcal{L}^{n}(O).
\end{equation}
The Krylov space of $O(t)$ is the linear span of operators constructed by repeated applications of $\mathcal{L}$ on $O$,
\begin{equation}
\label{span}
\mathcal{K}_{O} = span\{ O, \mathcal{L} O, \mathcal{L}^{2} O,...  \}.
\end{equation}
The operator $O$ is itself a vector $ \big| O \big) $ in the larger Hilbert space equipped with an inner product $\big (O \big| V \big) \equiv Tr(O^{\dagger} \rho_{1} V \rho_{2})$. 
Throughout this work, we will be working with Hilbert Schmidt inner product, i.e. $\rho_1 = \rho_2 = \mathbb{I}$. In this notation, the auto-correlation function takes the form
\begin{equation}
\label{auto_cf}
G(t) = \big (O \big| O(t) \big).
\end{equation}
The orthonormal basis for Krylov space can be constructed by applying the Lanczos algorithm. We fix the first Krylov operator $ \big | K_{0} \big ) = \big | O \big ) $ with $ \big (K_{0} \big| K_{0} \big) =1 $ and $b_{0} = 0$. Further operators can be constructed as
\begin{equation}
\big | K_{n} \big ) = \frac{1}{b_{n}}[ \mathcal{L} \big | K_{n-1} \big ) - b_{n-1} \big | K_{n-2} \big )~],
\end{equation} 
where the Lanczos coefficients $b_{n}$ are fixed such that Krylov basis operators are normalised to unity, i.e.,  $ \big (K_n \big| K_n \big) = 1 $. The algorithm is stopped whenever $b_{D_{\mathcal{K}}} = 0$ for some operator $ \big | K_{D_{\mathcal{K}}} \big ) $ which also fixes the dimension of the Krylov subsapce $ dim(\mathcal{K}_{O}) = D_{\mathcal{K}}$ which 
can in principle be $\infty$. The above algorithm suffers from numerical instabilities, which re-orthogonalisation algorithms can handle \cite{Rabinovici:2020ryf}. Having calculated the Krylov basis, one can define the $K$-complexity as
\begin{equation}
\label{kcom}
C_{K}(t) = \sum_{n}^{D_{\mathcal{K}}} n | \big( K_{n} \big | O(t) \big) |^{2} = \big(  O(t) \big |  \mathcal{K}  \big | O(t) \big)~,
\end{equation}
where $\mathcal{K}$ is the super operator which is diagonal in the Krylov basis $\mathcal{K} = diag(0,1,2,..,D_{\mathcal{K}})$. The $K$-complexity can 
also be defined as the average position of $\big| O(t) \big )$ in the Krylov basis.

In finite-dimensional Hilbert spaces, the notion of OQSL ~\cite{Carabba:2022itd,Hornedal:2023xpa} can be used to characterise the flow of operators under continuous evolution governed by the equation of motion. In this work, we will focus on the operators' Heisenberg evolution (see \cite{Hornedal:2023xpa} for other classes of flows). Following ~\cite{Carabba:2022itd,Hornedal:2023xpa}, the central quantity here is also the auto correlation function $ \big (O \big| O(t) \big) $ also known as the operator overlap. The derivation of the OQSL relies on the 
mapping between $d$ dimensional complex Hilbert space and $2d$ dimensional real vector space. This space is endowed with a Riemannian metric given to be the real 
part of $\big( . \big| . \big)$. This allows for the interpretation of $ \arccos(Re \big (O \big| U \big))  $ as the angle between two vectors in $\mathbb{R}^{2d} $. Since the norm of $O(t)$ is preserved under unitary evolution, the $ O(t) $ dynamics are contained on the $2d-1$ dimensional sphere with radius $|| O ||$ centred at the origin.

The OQSL for the complexity operator is (see Appendix(\ref{appen}) for the derivation)
\begin{equation}
\label{koqsl}
\tau_{ref}  = || \tilde{\mathcal{K}} || \frac{\arccos(\frac{ Re \big( \tilde{\mathcal{K}} (t) \big| \tilde{\mathcal{K}} \big ) }{|| \tilde{\mathcal{K}} ||^2})}{|| [ \mathcal{L} , \tilde{\mathcal{K}} ] ||}~,
\end{equation}
where $ \tilde{\mathcal{K}} (t) = e^{-i \mathcal{L} t} \tilde{\mathcal{K}}  e^{+i \mathcal{L} t}  $ and $  \tilde{\mathcal{K}} (t) = \mathcal{K} (t)- \big( \mathcal{K} (t) \big| \mathcal{I} \big ) \frac{\mathcal{I}}{|| \mathcal{I} ||^2}  $ with $\mathcal{V}(\tau)$ is replaced by $|| [ \mathcal{L} , \tilde{\mathcal{K}} ] ||$ velocity of complexity flow \cite{Hornedal:2023xpa}.

\subsection*{Relation to Quantum Chaos}
$K$-complexity and other Krylov space methods have been studied extensively in the context of quantum chaos in many body systems and field theories. It was conjectured in \cite{Parker:2018yvk} that in chaotic many-body systems in thermodynamic limit with local $O$, $b_n$ grow linearly with logarithmic correction in 1d systems. However, the exact relationship with other indicators of quantum chaos, such as level spacing distribution, OTOC, and spectral form factor, still needs to be determined. In finite many-body systems, the $b_n$ also have descent and plateau regimes after the initial linear growth.

The structure of eigenvalues of $\mathcal{H}$ is useful in highlighting some of the characteristics of the Krylov subspace. Level repulsion and Gaussian orthogonal ensemble (GOE) spectral statistics are the hallmarks of quantum chaos \cite{Bohigas:1983er}, while Poisson level spacing distributions are observed in non-chaotic models \cite{berry:tabor}. With 
the eigen decomposition $ \mathcal{H} = \sum_{m}^{d} E_{m} \big| E_{m} \big> \big< E_{m} \big|  $, if $ E_{m} = E_{n} $ with $m \neq n$ such scenarios are called resonances or degeneracies. Due to the level of repulsion, such conditions will be nearly improbable in chaotic systems. This places a constraint on the dimension of $Ker(\mathcal{L})$ \cite{Rabinovici:2020ryf} as the eigenvalues of the $\mathcal{L}$ are $E_{m} - E_{n}$. If $d$ is the dimension of $\mathcal{H}$, then $dim(Ker(\mathcal{L})) = d + ~number~of~resonances $ and the maximum possible dimension of Krylov space $D_{\mathcal{K}} = d^2 - dim(Ker(\mathcal{L})) + 1 $. It has also been seen numerically that the Lanczos sequence $b_{n}$ in systems with Poisson type level distribution have higher variances in $b_{n}$ ~\cite{Rabinovici:2021qqt,Hashimoto:2023swv}.

OQSL of the complexity operator $\mathcal{K}$ saturates when the operator $\mathcal{L}$ belongs to either the SU(2) or SL(2R) algebra ~\cite{Hornedal:2023xpa,Hornedal:2022pkc}. The nature of the OQSL of the complexity super operator $\hat{K}$ depends on the 2 resonance conditions, $E_{j} + E_{k} = E_{m} + E_{n}$ for $\{j,k \} \neq \{m,n \}$. This is evident by considering another super Liouvillian $ \mathbb{S} \equiv [\mathcal{L},.] $, the eigenvalues of $\mathbb{S}$ are given by $ E_{j} - E_{m} - ( E_{k} - E_{n} )$ where $ m,n,k,l \in \{ 1,2,...,D_{\mathcal{K}} \}$. As any general operator $A(t)$ evolving under the time evolution of $e^{-i \mathbb{S} t }$, can be decomposed into $A(t) = S + V(t)$, where the maximum dimension of $S$ is $D_{\mathcal{K}} +~
number~of~2~resonances$. Unlike the 1-resonance case, there can be many 2-resonances even in GOE spectral statistics. Consequently, in finite dimension many-body systems, 
the OQSL of the complexity operator will be sub-maximal as generic many-body systems have level repulsion and do not follow the SU(2) algebra.

Overall, the OQSL of the complexity operator is determined by the spectrum of $\mathbb{S}$ and the projection of $\mathcal{K}$ on the eigenbases of $\mathbb{S}$. However, due to the tensor product structure of the many-body Hamiltonian, it is worthwhile to study the Krylov basis operators $|K_{n})$ in a different operator basis that recognises the local structure of both $\mathcal{H}$ and $O$. This allows us to comment on the structure of the Krylov operators from an information-theoretic point of view. In the next section, we introduce the information-theoretic tools which are used to analyse the operators in the Krylov basis.

\section{Entanglement and Scrambling in Krylov space}
\label{secthree}

\begin{figure}
	\centering
    \includegraphics[width=0.5\textwidth]{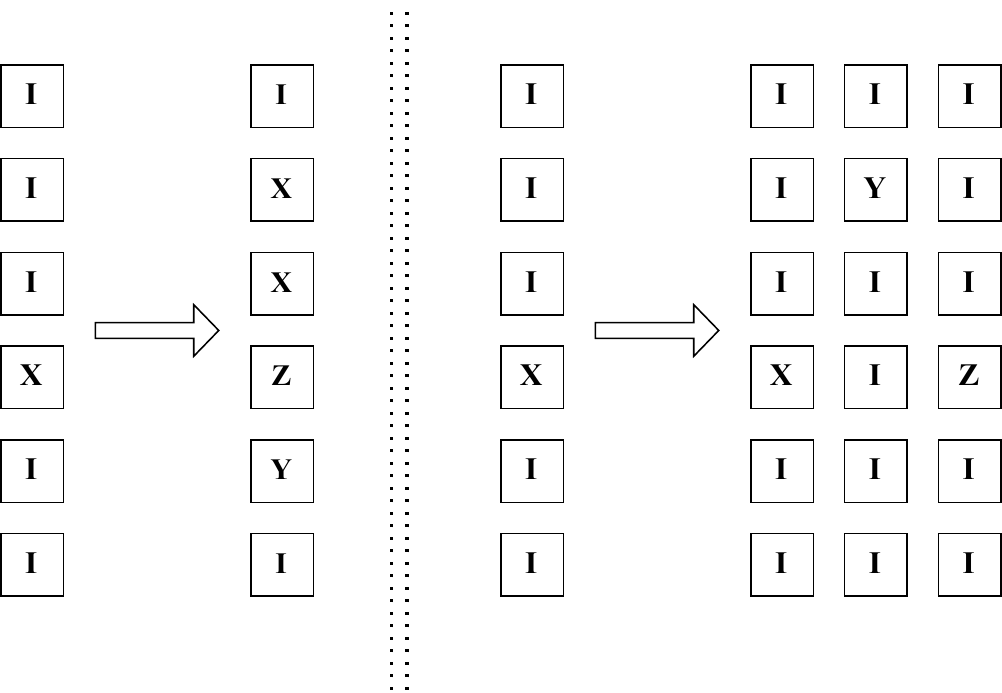}
	\caption{ Left - Entanglement Scrambling due to an increase in the size of the string of Pauli operators, Right- Magic Scrambling due to mapping of Pauli operators into superposition of multiple operators. \textcolor{black}{The arrow represents a unitary transformation.} Figure inspired from \cite{Garcia:2022abt} }
	\label{fig:ccram2}
\end{figure}

According to the resource theory of scrambling \cite{Garcia:2022abt,Hayden:2007cs,Shenker:2013pqa}, the mechanisms by which quantum information 
becomes scrambled can be categorized into two distinct classes: entanglement scrambling and Magic scrambling as shown in the representative 
Figure \ref{fig:ccram2}, which is inspired from \cite{Garcia:2022abt}. Here, ${\bf{\rm X}},{\bf{\rm Y}},{\bf{\rm Z}}$ denote the Pauli operators with ${\bf{\rm I}}$ 
being the identity. 
In entanglement scrambling, local Pauli operators, which initially affect only a tiny, localized part of the quantum system, evolve into Pauli operators of larger weight. This process spreads quantum information across a broader system region, increasing the entanglement and making it more challenging to extract the information without reversing the entire dynamics of the many-body system. Non-entangling unitaries do not increase the Pauli operators' weight; under conjugation, they take weight-1 Pauli operators to weight-1 Pauli operators.

On the other hand, Magic scrambling involves transforming strings of Pauli operators into a complex superposition of Pauli operators. The ``magic" of the quantum state refers to the non-stabilizer nature of quantum states that cannot be efficiently simulated by classical means. In this case, free unitary transformation (Clifford unitaries) on the Pauli operators preserves their structure, changing only the phase factor, but not creating a superposition of multiple Pauli operators. Magic scrambling of a non-clifford unitary is quantified by the distance between the unitary and the set of Clifford unitaries, identical to the resource theory of Magic \cite{Emerson:2013zse}.

These distinctions highlight how local quantum information spreads and is rendered inaccessible, contributing to our understanding of quantum chaos and the dynamics of complex quantum systems.

\subsection{Influence and Coherence}

We start with the generalised $n$-qubit Pauli group as $\mathcal{P}^{\otimes n}_{2} = \{P_{\vec{a}} : P_{\vec{a}} = \otimes^{n}_{i=1} P_{a_i} \}_{\vec{a} \in \mathbb{V}_{2}^{n}} $ with $P_{a_{i}} = X^{s_i} Z^{t_i}$ and $a_{i} = (s_{i},t_{i}) \in \mathbb{V}_{2} = \mathbb{Z}_{2} \otimes \mathbb{Z}_{2}$. $X$ and $Z$ are Pauli $Z$ and $X$ operators respectively. In this basis, it is natural to work with the inner product $\big< O_1 , O_2 \big > \equiv \frac{Tr(O_1^{\dagger} O_2)}{2^n}$ which is just the Hilbert Schmidt product normalised by the dimension of the Hilbert space. Any $n$-qubit operator $O$ can now be written as $O = \sum_{\vec{a} \in \mathbb{V}_{2}^{n}} c_{\vec{a}} P_{\vec{a}}$ with $c_{\vec{a}} = {\big< O , P_{\vec{a}} \big >} $. The normalisation condition $\sqrt{ \big< O , O \big >} = 1 $ implies $\sum_{\vec{a} \in \mathbb{V}_{2}^{n}} |c_{\vec{a}}|^2 = 1$.

 Now, the average size of an operator $O$ can be quantified by its influence \cite{Parker:2018yvk,Garcia:2022abt}. 
\begin{equation}
W(O) = \sum_{\vec{a} \in \mathbb{V}_{2}^n} |\vec{a}| |c_{\vec{a}}|^2
\end{equation}

Here, the size $|\vec{a}|$ is defined by the number of non-identity operators in the basis element $P_{\vec{a}}$. For instance, a Pauli basis element with size $|\vec{a}| = 2$ might include operators like $X \otimes I \otimes Z$ or $I \otimes Y \otimes X$. Influence is the average number (size) of non-Identity operator in the Pauli - basis expansion of $O$.
 Another valuable metric for assessing the spread of $O$ in the Pauli basis is the inverse participation ratio (IPR), given by $\frac{1}{\sum_{\vec{a}} |c_{\vec{a}}|^4}$, ~\cite{2010PhRvE..81c6206S,Anand:2020qhf,Bhattacharjee:2024yxj}. This ratio provides effective basis elements over which the operator $O(t)$ has significant support.
The IPR captures how coherently the operator $O$ is distributed across the Pauli operator basis. A low IPR indicates that $O$ is concentrated in the few basis elements, 
suggesting a less scrambled operator. Conversely, a high IPR signifies that $O$ is spread out over many basis elements, reflecting higher scrambling and complexity.
This metric serves as a measure of the coherence of $O$ with respect to the Pauli basis. By analyzing the IPR, we can gain insights into the degree to which $O$ has evolved, 
and how its components are distributed, providing a clearer picture of the operator's dynamics in the quantum system.

\subsection{Operator Entanglement}
Operator entanglement entropy (OpEE), which quantifies the entanglement of an operator \cite{2002PhRvA..66d4303W,2017PhRvB..95i4206Z, Alba:2019okd} is a key measure in studying quantum chaos and scrambling. To define the OpEE, we consider a bipartition of the spin chain into two subsystems, $\bold{A}$ and $\bold{B}$. Similar to the previous section, we can construct two separate bases for each of these subsystems, consisting of $(n_{\bold{A}} =3, n_{\bold{B}} =3)$ qubits, respectively. The basis vectors for these subsystems can be written as $A_{\vec{a}} = \otimes_{i=1}^{n_{\bold{A}}} P_{a_i}$, where $\vec{a} \in \mathbb{V}^{n_{\bold{A}}}_{2}$, and $B_{\vec{b}} = \otimes_{i=1}^{n_{\bold{B}}} P_{b_i}$, where $\vec{b} \in \mathbb{V}^{n_{\bold{B}}}_{2}$. Any operator $O$ can be decomposed uniquely $O = \sum_{\vec{a} \in \mathbb{V}^{n_{\bold{A}}}_{2}, \vec{b} \in \mathbb{V}^{n_{\bold{B}}}_{2} } O_{\vec{a},\vec{b}} A_{\vec{a}} \otimes B_{\vec{b}}$ where $O_{\vec{a},\vec{b}} = \big< A_{\vec{a}} \otimes B_{\vec{b}} \big| O \big>$. Now the OpEE of the normalised operator $O$ with $\sqrt{ \big< O , O \big>} = 1 $ can be defined as

\begin{equation}
S_{en} = -Tr(\rho^{\bold{A}}_{op} log(2, \rho^{\bold{A}}_{op}))~,
\end{equation}
where $(\rho^{\bold{A}}_{op})_{\vec{a},\vec{a'}} = \sum_{ \vec{b} \in \mathbb{V}^{n_{\bold{B}}}_{2} } O_{\vec{a},\vec{b}} O^{\dagger}_{\vec{b},\vec{a'}}  $.

\section{$K$-complexity and OQSL for Random Matrix theory and Many-body Hamiltonians}
\label{secfour}

\begin{figure}
\centering
\centering
\includegraphics[width=0.5\textwidth]{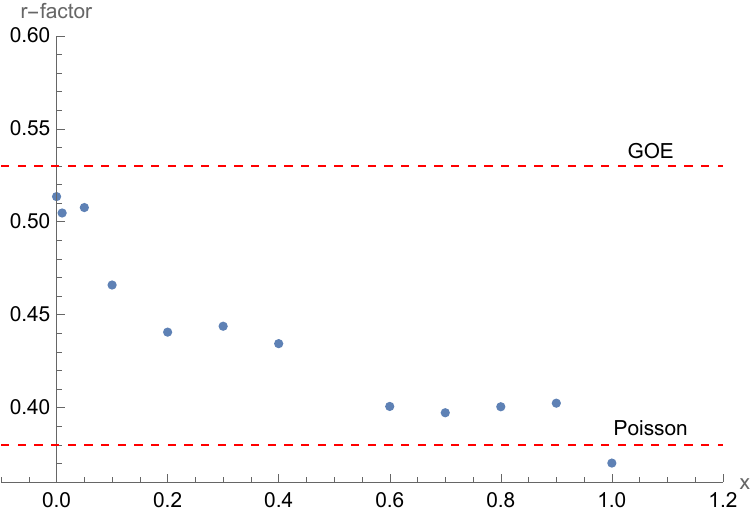}
\caption{\textcolor{black}{Behaviour of average gap ratios in the $D\times D$ (10 realizations) random/integrable matrix, for three consecutive levels as a function
of the parameter $x$.}}
\label{fig:rfac}
\end{figure}

In this section, we delve into the connection between the level spacing distribution and the complexity operator's OQSL. We aim to understand how the OQSL changes as we move from integrable systems to chaotic ones. Specifically, this crossover in level spacing distribution becomes evident when considering the level statistics of integrable matrices \cite{shastry:imt}. A matrix $\mathcal{H}(x) = x T + V $ is integrable if it has a commuting partner $ \tilde{\mathcal{H}}(x) = x \tilde{T} + \tilde{V} $ which is not a liner combination of $\mathcal{H} $ and $ I $ and there does not exist 
$\Omega$ such that $ [\Omega,\mathcal{H}] = [\Omega,\tilde{\mathcal{H}}] = 0 $. In this work, we will focus on type -1 integrable matrices which 
feature $ D-1 $ nontrivial commuting partners and use $D=64$. Any $ D \times D $ matrix can be parametrised as
\begin{equation}
\begin{aligned}
& {\mathcal{H}(x)_{i j}=x \gamma_i \gamma_j \frac{d_i-d_j}{e_i-e_j}, \quad i \neq j} \\
& {\mathcal{H}(x)_{j j}=d_j-x \sum_{k \neq j} \gamma_k^2 \frac{d_j-d_k}{e_j-e_k}}
\end{aligned}~,
\end{equation}
where $\gamma_{i}$ are sampled from the distribution $\delta(1 -  |\gamma|^{2})$, while $e_{i}$ and $d_{i}$ represent two sets of 
eigenvalues drawn independently from GOE ensembles. Integrable matrices exhibit a parameter dependence in their eigenvectors i.e., 
most eigenstates are localised in the eigenbasis of $V$. To study the effects of only level repulsion on OQSL, we need to remove this 
dependence by taking eigenvectors as Random vectors. At $x=1$, the level spacing follows Poisson statistics, transitioning to 
Wigner-Dyson statistics at $x=0$. We also plot the average gap ratio \cite{PhysRevB.75.155111} for three consecutive levels $r_i = min(\frac{\tilde{s}_i}{\tilde{s}_{i-1}},\frac{\tilde{s}_{i-1}}{\tilde{s}_i})$ with $\tilde{s_i} = E_{i} - E_{i-1}$ to quantify this behaviour Figure \ref{fig:rfac}. For GOE level spacing statistics gap ratio is $0.53$ and for Poisson level spacing statistics its 0.38. The $O$ operator is another random $D \times D$  matrix drawn from the GOE. This scenario has no notion 
of the locality of the operator $O$, making it an ideal setup for studying the dependence of OQSL and $K$-complexity solely on the level spacing.

\begin{figure}
\centering
\centering
\includegraphics[width=0.5\textwidth]{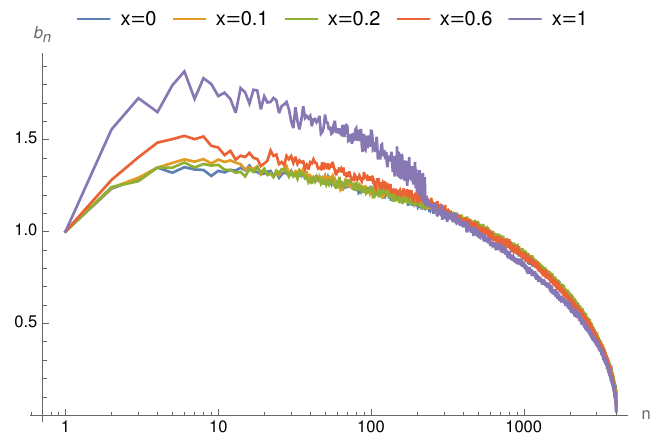}
\caption{\textcolor{black}{Behaviour of Lanczos coefficients $b_{n}$ for $D \times D$ random/integrable matrix (10
realizations) with parameter $x$.}}
\label{fig:rmt1}
\end{figure}

\begin{figure}
\centering
\centering
\includegraphics[width=0.5\textwidth]{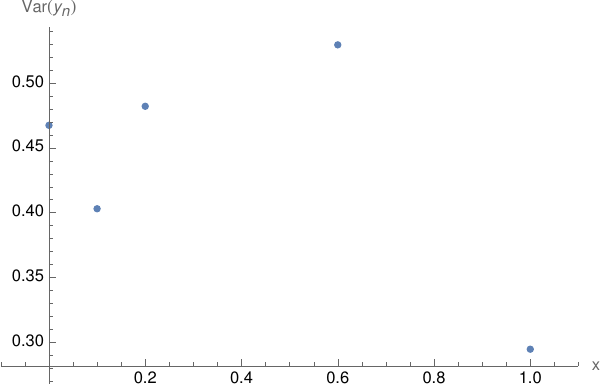}
\caption{\textcolor{black}{Behaviour of variance of $y_n= \frac{b_{2n-1}}{b_{2n}}$ for $D \times D $ random/integrable matrix (10 realisations) with parameter $x$.}}
\label{fig:varrmt}
\end{figure}

\begin{figure}
\centering
\centering
\includegraphics[width=0.5\textwidth]{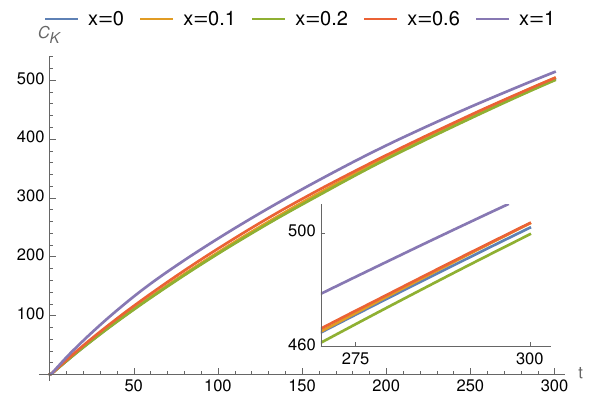}
\caption{\textcolor{black}{Behaviour of $C_{K}(t)$ for $D \times D$ random/integrable matrix (10
realizations) with parameter x.}}
\label{fig:rmtc}
\end{figure}
\begin{figure}
\centering
\centering
\includegraphics[width=0.5\textwidth]{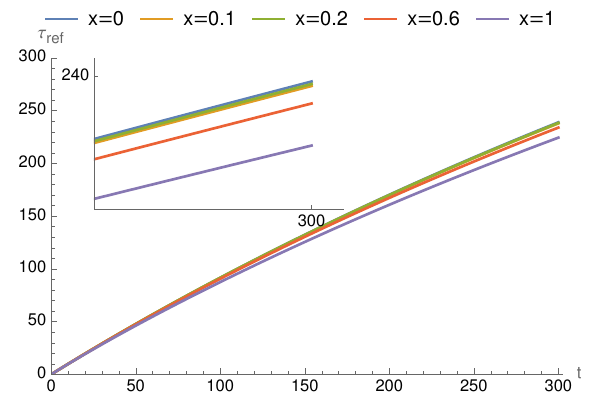}
\caption{\textcolor{black}{Behaviour of $OQSL$ as a function of $t$ for $D \times D$ random/integrable matrix (10
realizations) with parameter $x$.}}
\label{fig:rmt}
\end{figure}

The initial linear growth of the Lanczos coefficients $b_{n}$ (after normalizing the $b_{1}=1$) (Figure \ref{fig:rmt1}) decreases with the increase in repulsion level, which is followed by a brief plateau and descent. \textcolor{black}{To investigate the fluctuations in the behavior of $b_n$, we examine the variance of the variable $y_n \equiv \frac{b_{2n-1}}{b_{2n}}$ \cite{Hashimoto:2023swv} as a function of $x$, as shown in the Figure \ref{fig:varrmt}. The variance of $y_i$ initially increases, consistent with the transition from chaotic to integrable behavior, and then exhibits a decrease at the Poisson limit ($x = 1$). This unexpected sudden decrease may be due to the small matrix size ($D = 64$) or the limited number of realizations used in the analysis.} As the initial Lanczos coefficients determine the early behaviour of the $K$-complexity, the latter grows at faster rates in cases where level repulsion is minimal, see Figure \ref{fig:rmtc}. Since $O$ is not a local operator, the early linear growth of $K$-complexity cannot be associated with scrambling \cite{Rabinovici:2020ryf}. The complexity operator's OQSL (Figure \ref{fig:rmt}) first rises with increasing level repulsion, then slightly declines with a further increase in level repulsion. At late times (not shown), $C_{K}(t)$ and OQSL for different $x$ values saturates to similar values on nearly identical time scales.

Numerically, by analysing the Kernel of the super-operator $\mathbb{S} = [\mathcal{L},.]$ and the decomposition of $\mathcal{K}$ in the eigen-space of $\mathbb{S}$, we find that $\mathcal{I}$ is not the only stationary element under $e^{-i \mathcal{L} t}$ evolution. In other words, due to non-vanishing support of $|\mathcal{K})$  over $ker(\mathbb{S})$, the $OQSL$ is 
not tight with respect to bound (\ref{oqsl} for choice of $\mathcal{H}$ and $O$. This limit can be further refined by subtracting the projection of $|\mathcal{K})$ on $ker(\mathbb{S})$ from the $|\mathcal{K})$; for more details see \cite{Hornedal:2023xpa}. 

\begin{figure}
\centering
\centering
\includegraphics[width=0.5\textwidth]{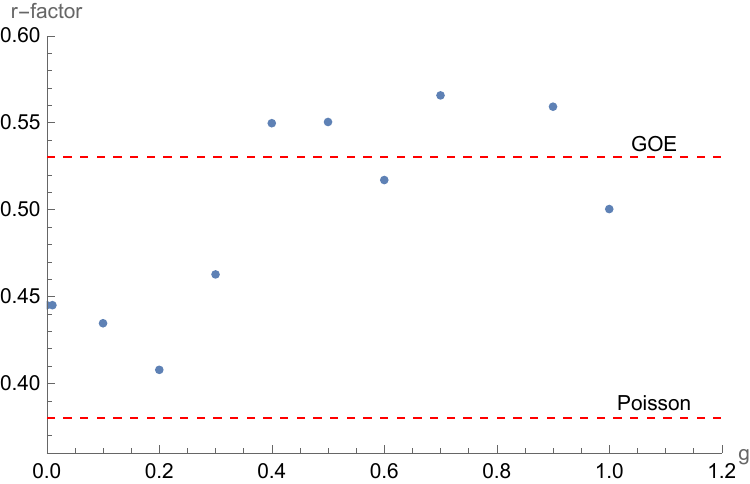}
\caption{\textcolor{black}{Behaviour of average gap ratios in the ANNNI model, for three consecutive levels with parameter $g$.}}
\label{fig:ranni11}
\end{figure}

\begin{figure}
	\centering
	\centering
	\includegraphics[width=0.5\textwidth]{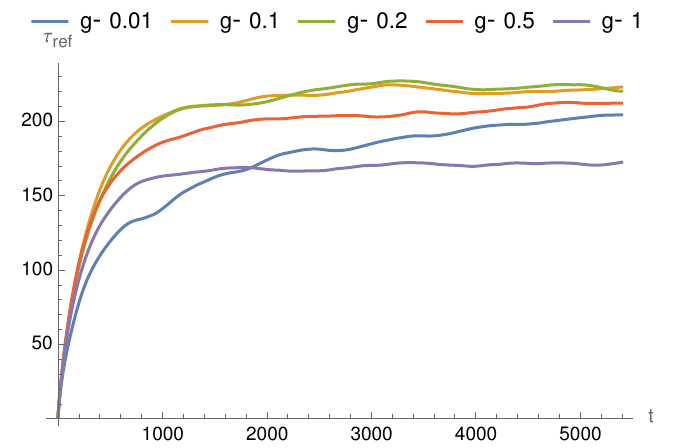}
	\caption{Behaviour of $OQSL$ as a function of $t$ for ANNNI model with integrability breaking parameter $g$.}
	\label{fig:anni}
\end{figure}

\begin{figure}
\centering
\centering
\includegraphics[width=0.5\textwidth]{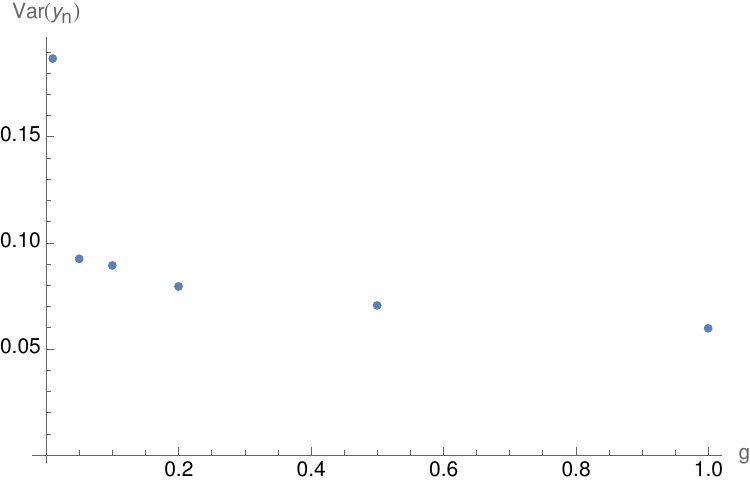}
\caption{\textcolor{black}{Behaviour of variance of $y_n =\frac{b_{2n-1}}{b_{2n}}$ with parameter $g$}}
\label{fig:varani}
\end{figure}

We recall that the GOE does not represent generic physical systems, since all energy levels interact. Hence, it is more prudent to study $OQSL$ and $K$-complexity in many-body 
systems from integrable to chaotic regimes. This work will consider the ANNNI \cite{anni:1,anni:2} Hamiltonian is a transverse field Ising chain with a non-integrable next-nearest-neighbour interaction term with open boundary conditions. The Hamiltonian for the model is 
\begin{equation}
\mathcal{H}(g,h) = - \sum_{i=1}^{L} (Z_{i} Z_{i+1} + h X_{i} + g Z_{i} Z_{i+2}),
\end{equation}
Since the ANNNI model is non-integrable for $|g|>0$, it can only be handled numerically through exact diagonalisation with $L = 6 $. \textcolor{black}{The average gap ratios for three consecutive levels shows GOE behaviour near $g=5$ (Figure \ref{fig:ranni11}).} We study the behaviour of $K$-complexity and $OQSL$ as we change the integrability breaking parameter $g>0$. Due to the tensor product structure of the Hamiltonian, the $OQSL$ and $K$-complexity will depend on the initial operator $O$ and its tensor product structure. This behaviour was absent in the previous case, as the eigenvectors of matrices drawn from the GOE are random.
\begin{figure}
	\centering
	\centering
	\includegraphics[width=0.5\textwidth]{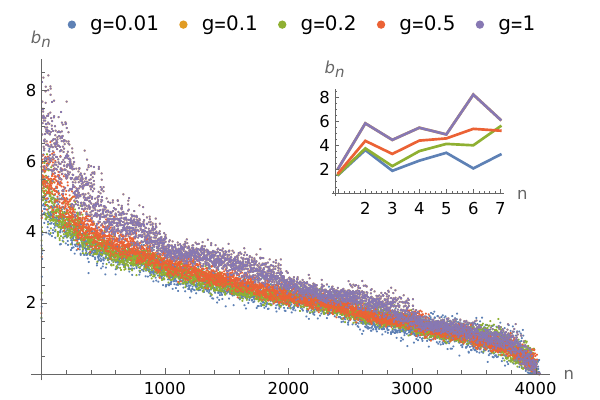}
	\caption{Behaviour of $b_{n}$ as a function of $n$ for ANNNI model with integrability breaking parameter $g$. Inset shows $b_{n}$ for small $n$.}
	\label{fig:anni1}
\end{figure}

\begin{figure}
	\centering
	\centering
	\includegraphics[width=0.5\textwidth]{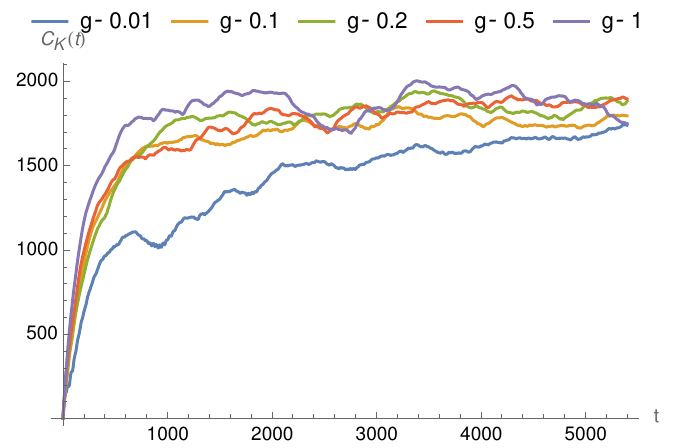}
	\caption{Behaviour of $C_{K}(t)$ for ANNNI model with integrability breaking parameter $g$.}
	\label{fig:annic}
\end{figure}

The OQSL (Figure \ref{fig:anni}) initially increases with the integrability breaking term $g$ and then decreases as $g$ increases. \textcolor{black}{To examine the irregular behavior of $b_n$ with respect to the integrability-breaking parameter, we plot the variance of the variable $y_n \equiv \frac{b_{2n-1}}{b_{2n}}$ as a function of $g$ Figure \ref{fig:varani}. The variance is observed to decrease as $g$ increases, which is expected due to the increased level repulsion, as also reflected in the behavior of the average gap ratio Figure \ref{fig:ranni11}.} A key difference in the behaviour of $b_{n}$ (Figure \ref{fig:anni1}) from the RMT case is the presence of an early rise in $b_{n}$ even for $g=1$, which is absent in the RMT $x=0$ case \cite{Bhattacharjee:2024yxj} although both of these lie in chaotic regimes. This is due to the local structure of $O$ which proportional to ($ 7X_{L/2} $ + $ 4 Z_{L/2} $  ) and $\mathcal{H}$. There will always be initial spreading of the local operator irrespective of the integrability of $\mathcal{H}$. In Figure \ref{fig:annic}, the complexity grows at faster rates for higher values of $g$, which is expected as the $b_{n}$ grows faster with $n$. This behaviour is in contrast with the RMT case. Similar to the RMT case,  this OQSL is not tight with respect to bound (\ref{oqsl}) for the ANNNI model, and we expect this to be the case for generic finite many-body systems. The late-time behaviour of both OQLS and K -complexity shows saturation at similar time scales.

The differences in the behaviours of OQSL and $K$-complexity between the RMT case and the ANNNI model are mainly because of the notion of locality. In the RMT case, there is no notion of locality due to the lack of the tensor product structure in both $\mathcal{H}(x)$ and  $O$. However, in the ANNNI case, both Hamiltonian and initial operator are local. This structure can be further explored by studying the properties of the eigenvectors of the complexity operator $\mathcal{K}$. In the next section, we see how the information-theoretic properties like entanglement and scrambling of these operators $\big| K_{n} \big)$ behave with the increase in the integrability breaking parameter.

\section{Entanglement and Scrambling in ANNNI model}
\label{secfive}

\begin{figure}
	\centering
	\centering
    \includegraphics[width=0.5\textwidth]{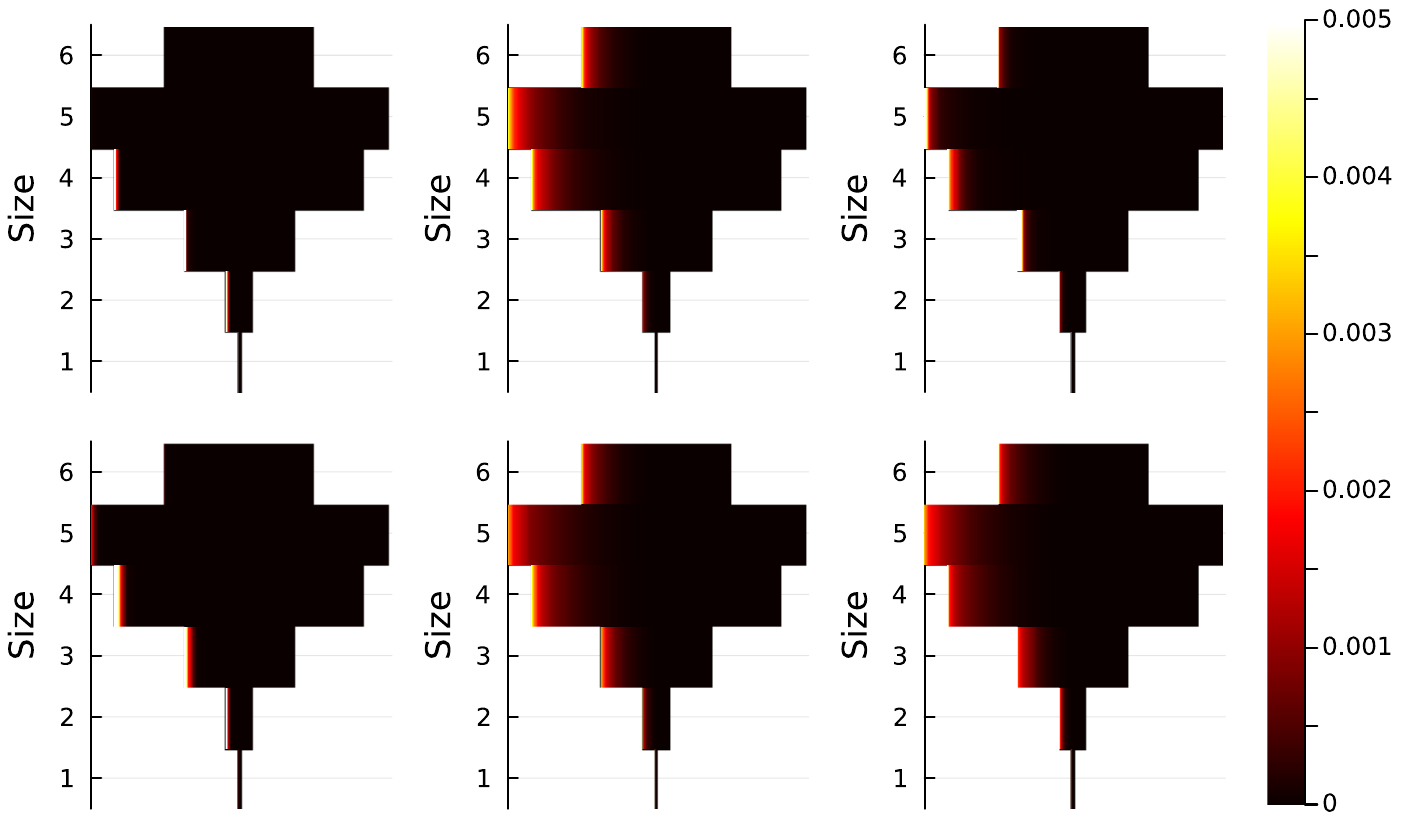}
	\caption{ Behaviour of $|c_{\vec{a}}|$ in the ANNNI model for different Krylov basis operators (left to right) $n = \{6,500,4000 \}$}  for $g = 0.01$ (top row) and $g = 0.5$ (bottom row). 
	\label{fig:matelements}
\end{figure}

To illustrate both entanglement and magic scrambling mechanisms in the ANNNI model, we plot the elements $|c_{\vec{a}}|$ via density plots, where each row represents the projection of the operator $O$ on operators of a fixed size. The elements in each row correspond to all the Pauli basis elements of a particular size $|\vec{a}|$. Each line in the horizontal strip along the X-axis in Figure \ref{fig:matelements} is a projection of operator $O$ on the Pauli operator of a particular size, depicted in the Y-axis. For example, the third strip of a graph shows the $|c_{\vec{a}}|$ of the projection of operator $O$ on the Pauli operators basis with exactly three non-identity operators. The density plot thus provides a visual representation of how the operator $O$ is projected onto these basis elements of varying sizes, highlighting the distribution of its components.
In this visualization, snapshots of Krylov basis operators show similar patterns for small and intermediate values of $n$ (e.g., $n = 6$ and $n = 500$). This similarity suggests that the nature of scrambling of $|K_{n})$ are relatively the same across these levels of $n$.
However, for a large  $n = 4000$, the pattern changes significantly depending on the coupling constant $g$. \textcolor{black}{In the case of $g=0.01$, the projection $|c_{\vec{a}}|$ onto Pauli operators of larger sizes is visibly smaller compared to the case of $g=0.5$. This behavior is also evident in Figure 12, where $IPR(K_{n})$ near $n=4000$ declines more rapidly for $g=0.01$ than for $g=0.5$. The color grading appears from left to right because the data for $|c_{\vec{a}}|$ is sorted in descending order from left to right.} This behaviour suggests that the operator exhibits less magic scrambling with smaller values of $g$, meaning that it has not diffused as broadly across the operator basis. This reduction in scrambling implies that the system's evolution allows for some unscrambling of initial quantum information. Overall, this density plot analysis shows how the complexity and distribution of the operator $O$ change under different levels of integrability breaking.
\begin{figure}
	\centering
	\centering
    \includegraphics[width=0.5\textwidth]{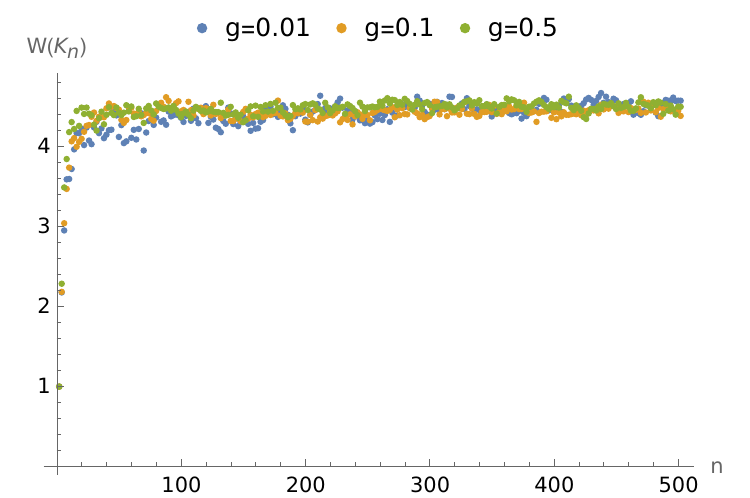}
	\caption{Influence $W(K_{n})$ as a function of $n$ in the ANNNI model for Krylov basis operators with integrability breaking parameter $g$.}
	\label{fig:inf}
\end{figure}

\begin{figure}
	\centering
	\centering
    \includegraphics[width=0.5\textwidth]{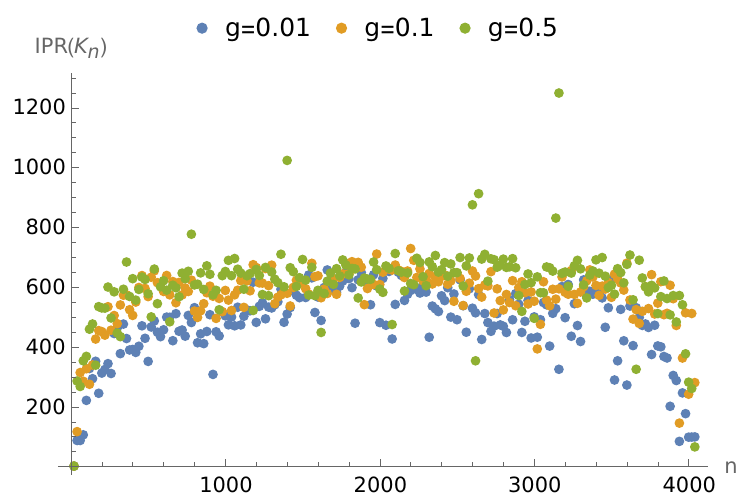}
	\caption{Inverse participation ratio $IPR(K_{n})$ in the ANNNI model as a function of $n$ of Krylov basis operators with integrability breaking parameter $g$.}
	\label{fig:iprk}
\end{figure}
\begin{figure}
	\centering
 
	\centering
    \includegraphics[width=0.5\textwidth]{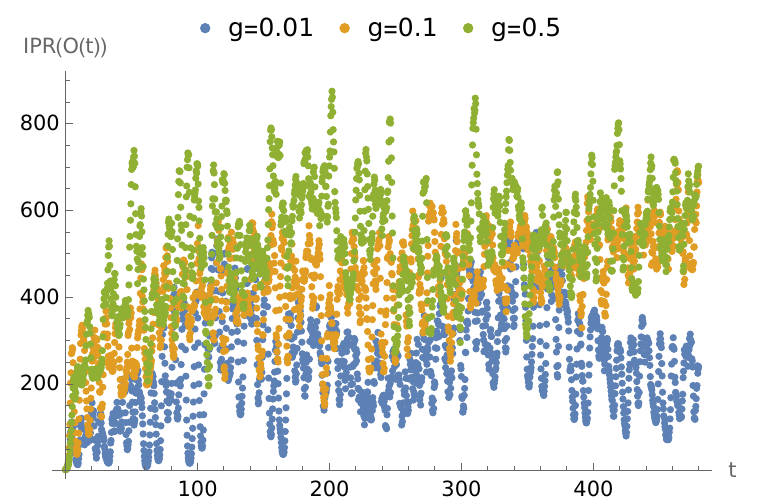}
	\caption{Time evolution of inverse participation ratio $IPR(O(t))$ of the  time evolved operator $O(t)$ with integrability breaking parameter $g$ in
	the ANNNI model.}
	\label{fig:ipr}
\end{figure}

For a unitary to function effectively as a magic scrambler, it must also be proficient in entanglement scrambling. Because the number of available operators to form a superposition increases dramatically as the size of the operator string increases. Therefore, examining the operator's influence and coherence is instructive for determining whether the dynamics generate entanglement and magic scrambling.

\subsection{Influence, Coherence and Entanglement in Krylov Operator space}

This section studies the influence and coherence of the time-evolved operator $O(t)$ and the Krylov basis operators $|K_{n} )$ generated during the Lanczos algorithm in the ANNNI model.
In Figure $\ref{fig:inf}$, we observe that the influence of the initial Krylov basis operators increases and then saturates for all subsequent Krylov basis operators, regardless of the value of $g$. It suggests that quantum chaos does not play a significant role in the influence dynamic of the operator.
In contrast, the inverse participation ratio of the Krylov basis operators $\big| K_{n} \big)$ (Figure \ref{fig:iprk}) and the time-evolved operator $\big| O(t) \big)$ (Figure \ref{fig:ipr}) exhibits distinct behaviours for different values of the integrability breaking parameter $g$. For a small $g = 0.01$, the IPR of $\big| K_{n} \big)$ peaks at central values of $n$, indicating that the operator is most coherent and de-localized in these regions, which suggests limited magic scrambling.
However, for larger values of $g = 0.1$ and $g = 0.5$, the IPR shows a plateau-like behaviour, implying that the operators  $|K_{n})$ has same localization in Pauli operator basis for $|K_{n})$ in the middle of Krylov operator basis. This distribution suggests more magic scrambling as the operator forms superpositions over a more extensive set of basis elements.
By analyzing both the influence and the IPR we can see how the system's dynamics facilitate entanglement and magic scrambling. This dual capability is essential for complex quantum systems to act as an effective scrambler of quantum information.
\begin{figure}
	\centering
 
	\centering
    \includegraphics[width=0.5\textwidth]{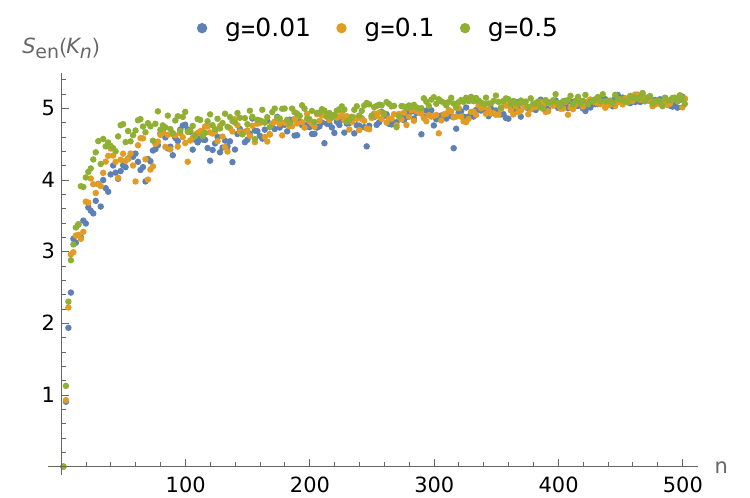}
	\caption{Operator entanglement entropy $S_{en}$ of Krylov basis operators ($K_{n}$) with integrability breaking  parameter $g$ in the ANNNI model.}
	\label{fig:opee}
\end{figure}
\begin{figure}
	\centering
 
	\centering
    \includegraphics[width=0.5\textwidth]{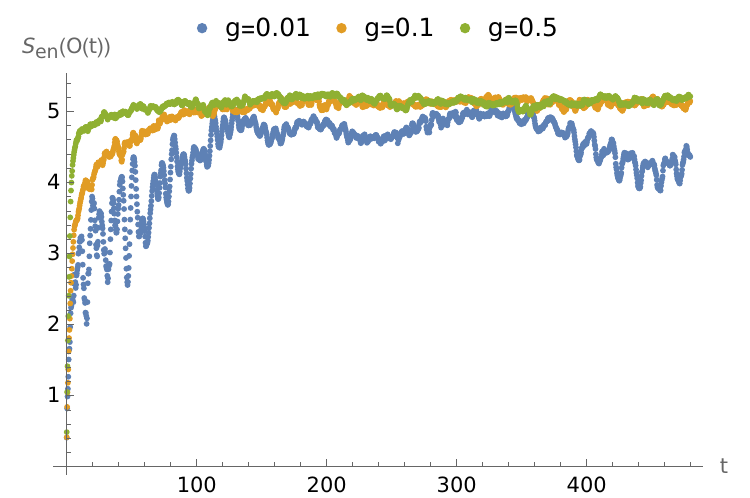}
	\caption{Operator entanglement entropy $S_{en}$ of the time evolved operator $O(t)$ with integrability breaking  parameter $g$.}
	\label{fig:opeee}
\end{figure}

Now we study the operator entanglement entropy of both $O(t)$ and Krylov basis operators $|K_{n})$.
In Figure $\ref{fig:opee}$, the operator entanglement entropy (OpEE) of the initial $\big| K_{n} \big)$ exhibits varying growth patterns for different values of $g$. As the integrability-breaking perturbation increases, the OpEE saturates more rapidly, increasing $n$. However, regardless of the values of $g$, the OpEE of higher $\big| K_{n} \big)$ remains saturated, indicating a stabilization of entanglement in Krylov basis operators. This behaviour contrasts with the OpEE of $O(t)$ Figure \ref{fig:opeee}, which never saturates even at late times for lower values of $g$.

\section{Discussions}
\label{secsix}

We explored the relationship between the operator quantum speed limit of the complexity operator $\mathcal{K}$ and level-repulsion in both random/integrable matrices and many-body systems. The OQSL depends on the spectrum of super Liouvillian $\mathbb{S}$ \cite{Hornedal:2022pkc} via 2 resonances. In the case of random/integrable matrices, we observed that the OQSL increases with the level repulsion, whereas the growth rate of $K$-complexity decreases. This indicates a correlation between the repulsion of energy levels and the speed at which the complexity operator can evolve. The decrease in the growth rate of $K$-complexity might be an artefact of taking eigenvectors of 
$\mathcal{H}(x)$ as random vectors for every value of $x$. 

In many-body systems, the behaviour is more complicated due to the presence of the tensor product structure. Initially, the OQSL increases with the integrability-breaking parameter $g$, suggesting that breaking integrability allows for faster evolution of the complexity operator. However, at larger values of $g$, the OQSL decreases, indicating a slowdown in the operator's evolution speed. 

Contrary to the random/integrable matrix case, the growth rate of $K$-complexity in many-body systems rises with increasing $g$. This suggests that as integrability is further broken, the complexity of the operators grows faster in many-body dynamics. 

For systems exhibiting chaotic dynamics, the Krylov operator $\big| K_{n} \big)$ remains a reliable representation of the coherence and entanglement properties of the time-evolved operator $O(t)$ even for moderate values of $n$, particularly at late times. This reliability comes from the complex nature of chaotic dynamics, which tend to spread $O(t)$ across all basis elements, leading to significant entanglement and scrambling.

The coherence of Krylov basis operators shows distinct behaviour for integrable dynamics, making it a better probe for chaos than opEE. The reliability is lacking when dealing with the entanglement dynamics. In integrable systems, the dynamics is more constrained, leading to limited spreading of $O(t)$ and reduced entanglement. As a result, the Krylov basis operators $\big| K_{n} \big)$ may fail to accurately capture the entanglement exhibited by the time-evolved operator $O(t)$, especially during late times. 

This distinction highlights the crucial role of chaotic dynamics in generating extensive entanglement and scrambling. In contrast, integrable dynamics impose constraints that can limit the ability of certain representations, such as the Krylov operator, to faithfully reflect the actual entanglement dynamics of the system.

\begin{center}
\bf{Acknowledgements}
\end{center}
We thank our anonymous referees for extremely useful comments that helped to improve a previous draft of this paper. 
We sincerely  thank Kunal Pal and Kuntal Pal for discussions. 
The work of TS is supported in part by the USV Chair Professor position at the Indian Institute of Technology, Kanpur.

\appendix
\section{Refined Operator Quantum Speed Limit}\label{appen}

The OQSL and the its refinements relies on
the geodesic distance between two operators $O$ and $ U $ lying on the sphere of radius $|| O ||$ is
\begin{equation}
\label{disop}
dist(O,U) = ||O|| \arccos(\frac{Re (O|U)}{||O||^{2}}).
\end{equation}

Now one can obtain the expression of the OQSL denoted by $\tau_{QSL}$ by the following process. First note that
\begin{equation}
\label{oqsl}
\tau = \frac{length(O(\tau))}{\frac{1}{\tau}length(O(\tau))} \geq
\frac{dist(O,O(\tau))}{\frac{1}{\tau} length(O(\tau))}~,
\end{equation}
where $length(O(\tau)) = \int_{0}^{\tau} || \mathcal{L}(t) O(t)|| dt $ and $ \mathcal{V}(\tau) = \frac{1}{\tau} \int_{0}^{\tau} || \mathcal{L}(t) O(t)|| dt $ is average speed of the evolution. Noting that any curve traced by $O(t)$ has to be greater than or equal to geodesic distance, the speed limit can be obtained as
\begin{equation}
\tau \geq \tau_{QSL} = \frac{ \sqrt{G(0)} \arccos(\frac{Re(G(\tau))}{G(0)}) }{\mathcal{V}(\tau)}~.
\end{equation}
One can obtain a refinement to the above OQSL \cite{Hornedal:2023xpa} by separating the part of $O(t)$ which is stationary under unitary evolution $e^{-i \mathcal{L} t}$ as $O(t) = S + U(t)$ where $\big( S \big| U(t) \big ) = 0 $ throughout the evolution. Hence, the refined $OQSL$ is
\begin{equation}
\label{rOQSL}
t_{ref}(\tau) = \frac{\sqrt{G(0) - ||S||^2} \arccos(\frac{Re(G(\tau) - ||S||^2)}{G(0) - ||S||^2}) }{\mathcal{V}(\tau)}.
\end{equation}
Having defined the refined OQSL and since $[\mathcal{I},\mathcal{L}] = 0$ (here $\mathcal{I}$ is the identity operator in same Hilbert space as $\mathcal{L}$), 
the OQSL for the complexity operator after this refinement is  
\begin{equation}
\label{koqslder}
\tau_{ref}  = || \tilde{\mathcal{K}} || \frac{\arccos(\frac{ Re \big( \tilde{\mathcal{K}} (t) \big| \tilde{\mathcal{K}} \big ) }{|| \tilde{\mathcal{K}} ||^2})}{|| [ \mathcal{L} , \tilde{\mathcal{K}} ] ||}~,
\end{equation}
where $ \tilde{\mathcal{K}} (t) = e^{-i \mathcal{L} t} \tilde{\mathcal{K}}  e^{+i \mathcal{L} t}  $ and $  \tilde{\mathcal{K}} (t) = \mathcal{K} (t)- \big( \mathcal{K} (t) \big| \mathcal{I} \big ) \frac{\mathcal{I}}{|| \mathcal{I} ||^2}  $ with $\mathcal{V}(\tau)$ is replaced by $|| [ \mathcal{L} , \tilde{\mathcal{K}} ] ||$ velocity of complexity flow \cite{Hornedal:2023xpa}.

\bibliography{biblio}
\end{document}